\documentclass[aps,prb,twocolumn,superscriptaddress]{revtex4}
\usepackage{color}
\usepackage{amsmath}
\usepackage{graphicx}
\usepackage{tabularx}
\usepackage{textcomp}
\usepackage{gensymb}
\usepackage{keyval}
\usepackage{multirow}

\begin{document}
\title{Direct observation of a Fermi liquid-like normal state in an iron-pnictide superconductor.}
\author{Alona Tytarenko}
\author{Yingkai Huang}
\author{Anne de Visser}
\affiliation{van der Waals - Zeeman institute, University of Amsterdam, 1018 XL Amsterdam, the Netherlands}
\author{Steve Johnston}
\affiliation{Department of Physics and Astronomy, University of Tennessee, Knoxville, U.S.A.}
\author{Erik van Heumen}
\email{e.vanheumen@uva.nl}
\affiliation{van der Waals - Zeeman institute, University of Amsterdam, 1018 XL Amsterdam, the Netherlands}

\date{\today}
\begin{abstract}
There are two prerequisites for understanding high-temperature (high-T$_c$) superconductivity: identifying the pairing interaction and a correct description of the normal state from which superconductivity emerges. The nature of the normal state of iron-pnictide superconductors, and the role played by correlations arising from partially screened interactions, are still under debate. Here we show that the normal state of carefully annealed electron-doped BaFe$_{2-x}$Co$_{x}$As$_2$ at low temperatures has all the hallmark properties of a local Fermi liquid, with a more incoherent state emerging at elevated temperatures, an identification made possible using bulk-sensitive optical spectroscopy with high frequency and temperature resolution. The frequency dependent scattering rate extracted from the optical conductivity deviates from the expected scaling $M_{2}(\omega,T)\propto(\hbar\omega)^{2}+(p\pi k_{B}T)^{2}$ with $p\approx$ 1.47 rather than $p$ = 2, indicative of the presence of residual elastic resonant scattering. Excellent agreement between the experimental results and theoretical modeling allows us to extract the characteristic Fermi liquid scale $T_{0}\approx$ 1700 K. Our results show that the electron-doped iron-pnictides should be regarded as weakly correlated Fermi liquids with a weak mass enhancement resulting from residual electron-electron scattering from thermally excited quasi-particles.  
\end{abstract}
\maketitle

\begin{figure*}
\includegraphics[ width = 17.6 cm]{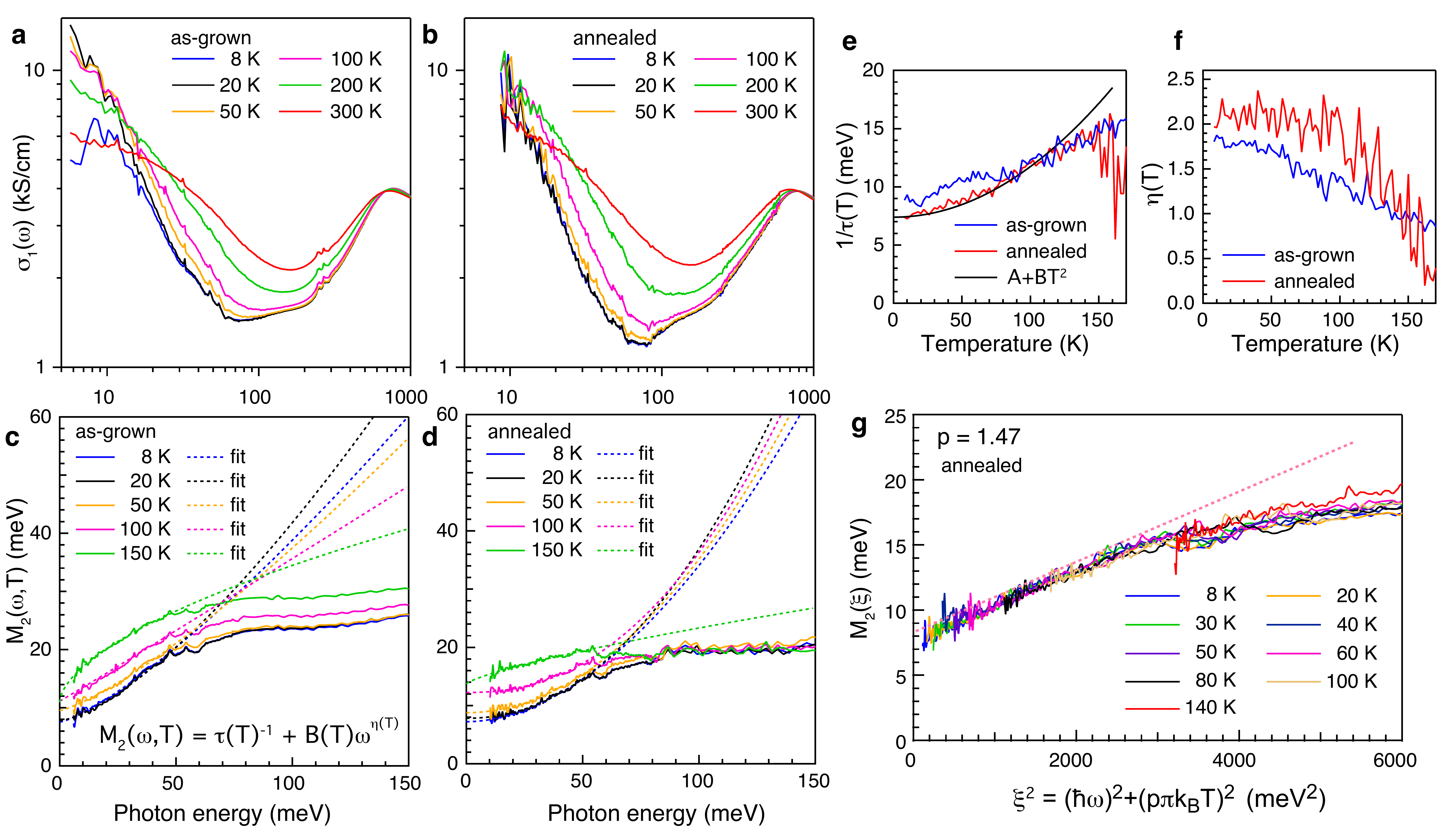}
 \caption{
 {\bf Experimental determination of Fermi liquid behaviour in electron doped BaFe$_2$As$_2$}.
 {\bf a}, {\bf b}, Comparison of the real part of the optical conductivity $\sigma_1(\omega)$ for as-grown (T$_{c}\approx$ 18 K) and annealed (T$_{c}\approx$ 25 K) BaFe$_{1.8}$Co$_{0.2}$As$_2$ at selected temperatures. The most significant, annealing induced change is a reduction of a broad incoherent background that is most clearly seen by the deeper minimum around 70 meV separating the free charge and inter-band optical conductivity. {\bf c}, {\bf d}, Imaginary part of the memory function, $M_2(\omega,T)$, revealing the difference in free charge response for as-grown and annealed crystals. The memory function is obtained by subtracting the full interband response as discussed in the text and SOM\cite{som_annpnic}. Dashed curves indicate fits made using the fitting function indicated in panel {\bf c}. {\bf e}, Temperature dependence of the static scattering rate $1/\tau(T)$ obtained from the fits in panels {\bf c}, {\bf d}. For the annealed crystal $1/\tau(T)$ displays a T$^2$ behaviour below $T \sim$ 100 K as indicated by the fit. {\bf f}, Temperature dependence of the exponent, $\eta(T)$,  extracted from the fits in panel {\bf c}, {\bf d}. The exponent for the annealed crystal shows $\omega^2$ dependence in the same temperature range where $1/\tau(T)$ has a $T^2$ temperature dependence. At higher temperatures a clear deviation from Fermi liquid behaviour is found. {\bf g}, Scaling collapse obtained by plotting $M_2(\omega,T)$ as $M_2(\xi)$, where $\xi^2 = (\hbar\omega)^2 + (1.47\pi k_BT)^2$. Above $\xi^2 \approx 2500$ meV$^2$ the scaling deviates from the universal Fermi liquid behaviour which is indicated by the dashed pink line.
 }
\end{figure*}

Strong electronic correlations and Mott physics have played an important role in shaping our understanding of high-$T_c$ superconductivity (HTSC).\cite{lee:2006ba} With the discovery of the iron-pnictide family of HTSCs a new playground to study correlation effects has emerged.\cite{Georges:2013ju} Unlike the cuprate HTSC, the pnictides are properly classified as moderately correlated semi-metals \cite{Qazilbash:2009ca}. By studying their normal state properties a new picture has started to emerge \cite{Haule:2009be} where intra-atomic exchange processes (Hund's coupling) govern the degree of correlation effects. In the resulting ``Hund's metal'' state\cite{Yin:2011ge}, Hund's coupling reduces the propensity towards a strongly correlated Mott insulating state, while simultaneously reducing the coherence temperature below which Fermi liquid (FL) properties emerge. A strong dependence of the nature of this Hund's metal state on orbital filling has been found, providing a natural explanation for the differences between hole- and electron-doped pnictides.\cite{Georges:2013ju, deMedici:2014wk} Recently, Werner \textit{et al.} showed\cite{werner:2012cs} that the combined effect of dynamic screening (manifested through a single particle self-energy, $\Sigma(\omega,T)$) and orbital occupancy results in a Fermi-liquid like state in electron-doped pnictides, while a spin-freezing transition separates an incoherent metal regime from the FL regime in hole-doped materials  (for a more extensive review of the role of Hund's coupling in the iron-pnictides, see ref. [2]). A clear experimental identification of both these regimes is currently lacking. Here, we provide direct experimental confirmation of the Fermi liquid state in the electron-doped case. 

Optical spectroscopy is a powerful tool to probe self-energy effects \cite{Basov:2010un} as a function of frequency and temperature simultaneously. The complex-valued free charge optical conductivity\cite{goetze:1972as} can be written as 
\begin{equation}\label{EDM}
\sigma(\omega,T)=
\sigma_1(\omega,T) + i\sigma_2(\omega,T) = 
\frac{i\omega_{p}^{2}}{4\pi}\frac{1}{\omega+M(\omega,T)}, 
\end{equation}
where $\omega_{p}^{2}=ne^{2}/m$ is the plasma frequency and $M(\omega,T) = M_1(\omega,T) + iM_2(\omega,T)$ is the complex memory function. For a simple Drude metal $M(\omega,T) = i\Gamma_D$ is frequency independent, while interactions beyond simple impurity scattering introduce a frequency and temperature dependence. In the latter case Eq. (\ref{EDM}) is referred to as the ``Extended Drude model''. The single particle self-energy $\Sigma(\omega,T)$ thus manifests itself in the free charge carrier response, appearing as a deviation from a classical Drude response. For a local FL with a momentum independent interaction between electrons, $\Sigma(\omega,T)$ and consequently $M(\omega,T)$ (see Methods) follow a universal quadratic dependence on both energy and temperature,\cite{Berthod:2013qq,Maslov:2012ba,Mirzaei:2013wf,Stricker:2014cx}
\begin{equation}\label{MFL}
M_{2}(\omega,T) = \frac{2}{3\pi k_{B}T_{0}}\left[(\hbar\omega)^{2}+(p\pi k_{B}T)^{2} \right], 
\end{equation}
where $k_BT_{0}$ is an overall energy scale characterising the correlation strength and $p$ is a non-universal constant. For a local FL one expects $p=2$, however deviations arise in the presence of additional elastic resonant scattering channels.\cite{Maslov:2012ba} To date the only known example with $p=2$ is Sr$_{2}$RuO$_{4}$\cite{Stricker:2014cx}, while $p\ne2$ has been reported for several correlated materials\cite{Mirzaei:2013wf,yang:2006tr,Katsufuji:1999cy, nagel:2012le, Dressel:2011bp}. When applied to the iron-pnictide superconductors, the accurate determination of $M(\omega,T)$ is hampered by the presence of low-lying interband transitions. In the following we first show that $M(\omega,T)$ extracted for carefully annealed BaFe$_{2-x}$Co$_{x}$As$_{2}$ single crystals indeed displays the characteristic $\omega,T$-scaling predicted by Eq. (\ref{MFL}). We then introduce an analysis of the complex optical conductivity that represents a direct confirmation of the Fermi liquid normal state of these electron doped iron-pnictides . This is made possible by the 2 K temperature resolution in our experiments, which allows us to compare the detailed frequency $and$ temperature dependence with similar resolution. 

\section{Results}
\begin{figure*}
\includegraphics[ width = 15 cm]{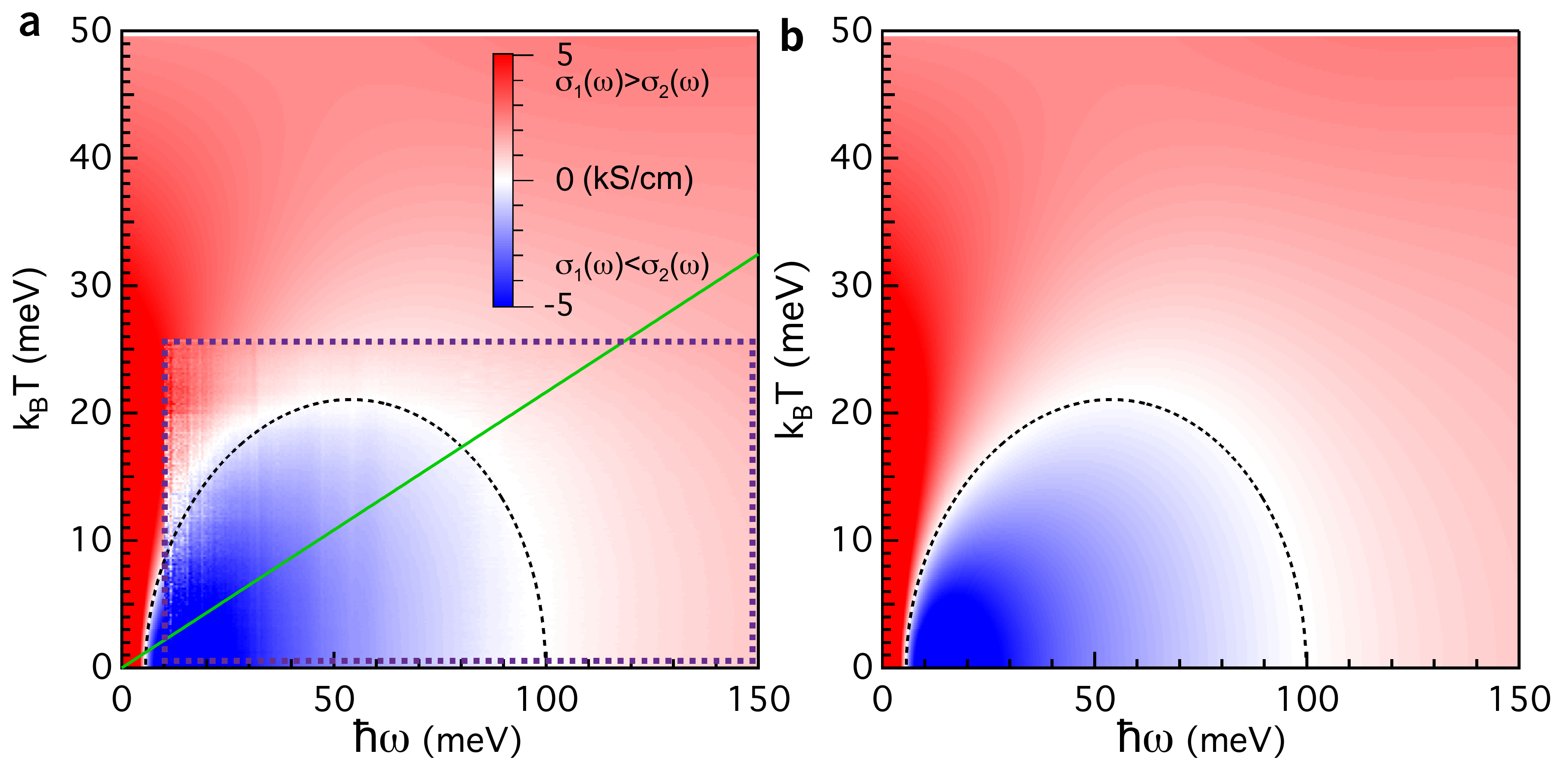}
\caption{\textbf{Fermi liquid behaviour of the optical conductivity: theory vs. experiment.} \textbf{a}, ($\omega$,T) dependence of $\Delta\sigma(\omega,T)\equiv\sigma_{1}(\omega,T)-\sigma_{2}(\omega,T)$. The experimental data is bounded by the purple dashed box, while the background image is the same as in panel \textbf{b}. Colour is used to indicate the magnitude of $\Delta\sigma(\omega,T)$, with red indicating the dissipative regime ($\sigma_{1}(\omega,T)>\sigma_{2}(\omega,T)$) and blue indicating the inductive regime ($\sigma_{1}(\omega,T)<\sigma_{2}(\omega,T)$). The colour scale is chosen such that the boundary between these two regimes, where $\sigma_{1}(\omega,T)\approx\sigma_{2}(\omega,T)$, or
 $\Delta\sigma(\omega,T)\approx$ 0, appears as white. This dome of zeroes can be reproduced using the approximate expression $T_{\infty}(\omega)$. The green line indicates the crossover temperature 1.47$\pi k_{B}T=\hbar\omega$, below which Fermi liquid behaviour can be
 expected. \textbf{b}, Same as in panel \textbf{a}, but calculated from the Allen-Kubo formula for the optical conductivity using a Fermi liquid self-energy with parameters derived from the experimental data of Fig. 1. The dashed semi-circle is the same as in panel \textbf{a}.}
\end{figure*}

The in-plane optical conductivity of as-grown and annealed BaFe$_{1.8}$Co$_{0.2}$As$_{2}$ (see Methods) are shown in Fig. 1{\bf a} and 1{\bf b}, respectively. After annealing we observe a decrease in the depth of the minimum around 70 meV separating the free-charge and interband optical conductivity. This arises from a reduction of a broad incoherent response associated with high energy interband processes, rather than from changes in the free-charge response\cite{som_annpnic}. A spectral weight analysis\cite{som_annpnic} for both crystals gives $\omega_{p}\approx$ 1.4 eV and a contribution of interband transitions to the low energy dielectric constant $\varepsilon_{\infty,IR}\approx$ 100. These similarities indicate that annealing does not significantly change the overall electronic structure (such as a chemical potential shift) or high-energy optical properties. 

\subsection{Experimental signatures of the Fermi liquid state.\\}
Subtle changes in the free charge carrier response are more easily analysed in terms of equations (\ref{EDM}) and (\ref{MFL}), but the extended Drude model analysis assumes that interband transitions do not contribute to the optical conductivity in the energy range of interest. The multi-band nature of the pnictides complicates the extraction of $M(\omega,T)$ since inter-band processes have a significant contribution to the optical conductivity\cite{VanHeumen:2010td,Benfatto:2011gn,Marsik:2013iv,Calderon:2014tfa}.  In the Supplementary material\cite{som_annpnic} (SOM) we describe the procedure used to extract the memory functions and its range of validity. We find that even though the determination of the memory functions comes with uncertainty at higher energies, at low energies ($\hbar\omega\leq$ 50 meV) interband transitions only weakly affect the frequency dependence. In the following we subtract the full frequency dependence of the interband response as outlined in the supplementary materials\cite{som_annpnic}; however, we note that our conclusions remain the same when alternative methods for accounting for the interband transitions are applied. 

The frequency and temperature dependence of the imaginary part of the memory function $M_{2}(\omega,T)$, shown in Fig. 1{\bf c} and 1{\bf d} for the as-grown and annealed crystal, respectively, indicates the presence of residual interactions beyond a classical Drude response. We fit both datasets with a power-law form $M_{2}(\omega,T)=1/\tau(0,T)+B(T)\omega^{\eta(T)}$, where $1/\tau(0,T)$ is the zero-frequency scattering rate and $\eta(T)$ = 2 is expected for a FL. These parameters are determined independently at each temperature. The temperature dependence of $1/\tau(0,T)$ and $\eta(T)$ are displayed in Fig. 1{\bf e,f} for both the as-grown and annealed crystal. We find that the annealed crystal displays characteristic FL behaviour with $\frac{1}{\tau(T)}\sim T^{2}$ (Fig.  1{\bf e}) and $\eta(T) \sim 2$ (Fig. 1{\bf f}) over a large range of energy (10 meV $\leq\hbar\omega\leq$ 50 meV) and temperature (8 K $\leq T\leq$ 100 K). We further find that the prefactor $B(T)$ is temperature independent in the same temperature range as is expected from Eq. (\ref{MFL}) (see SOM\cite{som_annpnic}). The as-grown crystal on the other hand does not display FL behaviour. Instead, the zero frequency scattering rate follows a more linear temperature dependence, while the frequency exponent $\eta(T) < 2$. Given the approximate $T^{2}$ and $\omega^{2}$ dependence of the memory function apparent in Fig. 1{\bf e,f}, we test whether the scaling form of Eq. \ref{MFL} applies to the annealed crystal. Fig. 1{\bf g} demonstrates that $M_{2}(\omega,T)$ indeed follows a universal FL scaling as function of the scaling variable $\xi^{2}=(\hbar\omega)^{2}+(p\pi k_{B}T)^{2}$, with $p\approx$ 1.47 (see SOM\cite{som_annpnic}). 

We highlight three deviations from universal FL behaviour that can be discerned in Fig. 1{\bf g}. First, universal FL behaviour disappears above 100 K. Second, for $\xi^{2}\geq$ 2500 meV$^{2}$, $M_{2}(\xi)$ changes slope, as indicated by the dashed pink line, signalling a crossover to a nearly energy independent $M_{2}(\omega,T)$ for $\hbar\omega\geq$ 50 meV (Fig. 1{\bf d}). Third, $p$ = 1.47 rather than 2, indicating that an additional elastic contribution is present beyond residual electron-electron scattering. We note that the precise value of $p$ determined by collapsing the data on a universal curve comes with some uncertainty as it depends on the assumed strength and frequency dependence of the interband contribution (see SOM\cite{som_annpnic}). 

\subsection{Fermi liquid signatures in the optical conductivity.\\}
Our analysis provides compelling evidence that the normal state of BaFe$_{1.8}$Co$_{0.2}$As$_{2}$ below 100 K is properly classified as a FL. We emphasise that the specific method of accounting for interband processes does not alter the conclusion that the low frequency and temperature dependence of $M_{2}(\omega,T)$ follows $\omega^{2}$ and $T^{2}$ scaling\cite{som_annpnic}.  In contrast, the same analysis applied to the as-grown crystal does not show such clear signatures of FL behaviour, despite its similar plasma frequency and high-energy optical properties. Nevertheless, the determination of the parameters characterising the Fermi liquid state using the extended Drude analysis remains sensitive to the choice for the interband contribution. To fortify our conclusions, and to determine the characteristic properties of the Fermi liquid state more accurately, we now turn our attention to an analysis of the complex optical conductivity, which provides a more direct comparison between theory and experiment and does not require a model specific choice for the interband processes.

\begin{figure*}
\includegraphics[ width = 15 cm]{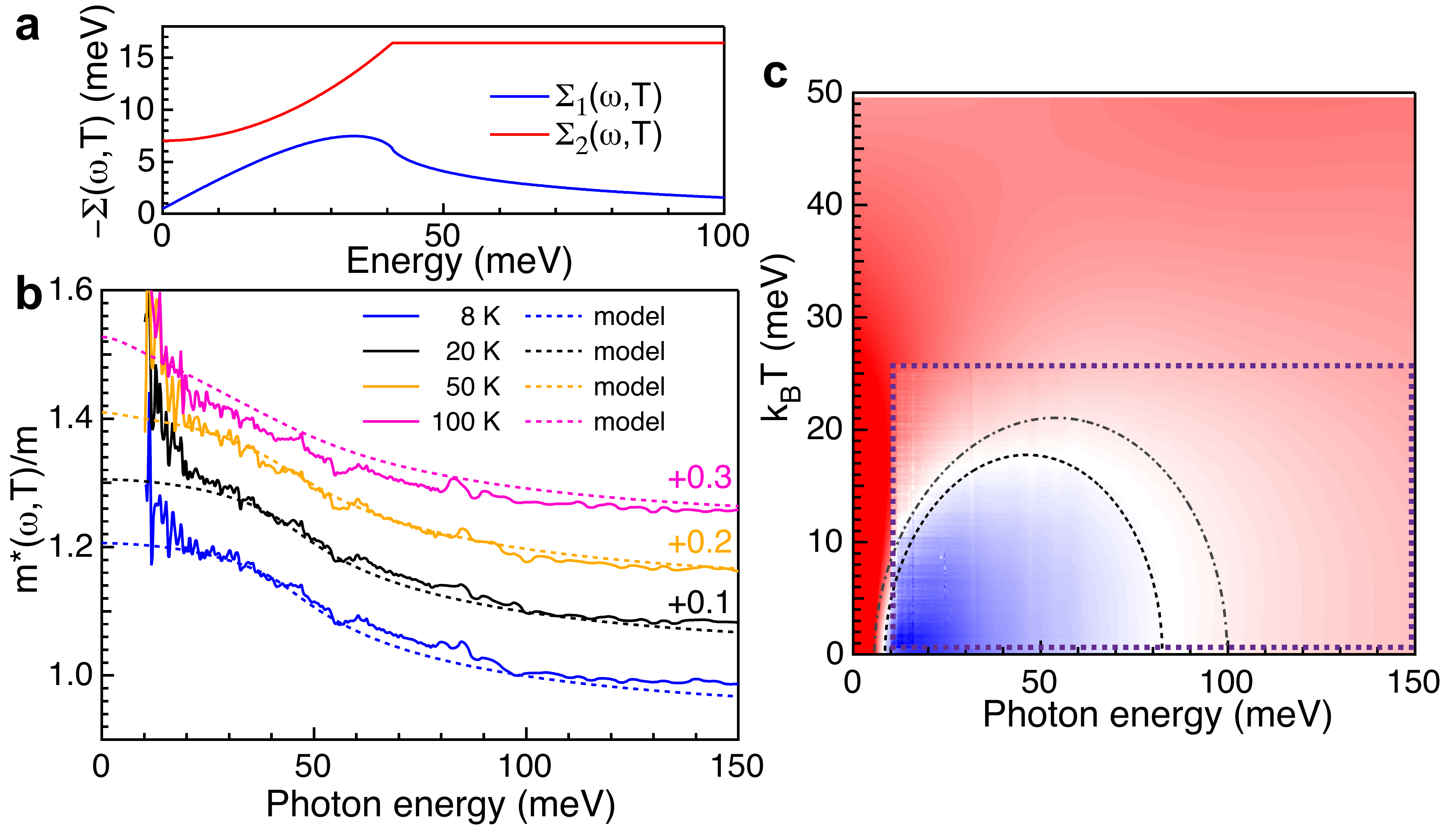}
\caption{\textbf{Self-energy, mass enhancement and Fermi liquid properties of as-grown BaFe$_{1.8}$Co$_{0.2}$As$_{2}$.} {\bf a}, single particle self-energy, $\Sigma(\omega,8 K)$, extracted from the optical data. The cutoff energy in $\Sigma_{2}(\omega,T)$ at $\omega_{c}$ = 41 meV introduces a finite slope of $\Sigma_{1}(\omega,T)$ at lower energy and a corresponding mass enhancement. \textbf{b}, energy and temperature dependence of the effective mass of the annealed crystal corresponding to the optical scattering rate of Fig. 1{\bf d}. The experimental mass enhancement is shown with solid lines, while the mass enhancement calculated from the Allen-Kubo formula is shown as dashed lines. The effect of the cutoff energy $\omega_{c}$ results in a mass enhancement $m^{*}/m(\omega\rightarrow 0)\approx$ 1.2. Note that the data and fits are offset from their actual value with increments of 0.1 at successive temperatures above the 8 K curves. \textbf{c},  ($\omega$,T) dependence of $\Delta\sigma(\omega,T)$ for the as-grown crystal. The dome of zero-crossings is smaller compared to the annealed crystal. This difference is highlighted by the dashed (as-grown) and dashed-dotted (annealed) semi-circles calculated from $T_{\infty}(\omega)$. The dashed semi circle is calculated using the same parameters as in Fig. 2 except for $\Gamma_{0}\approx$ 8 meV and $p\approx$ 1.34.}
\end{figure*}

Berthod \textit{et al.} showed\cite{Berthod:2013qq} that in a local FL a dome is defined by the locus of points where $\sigma_1(\omega,T) = \sigma_2(\omega,T)$, which bounds a 'thermal' regime in which FL behaviour emerges. Zero crossings signalling the presence of a dome have been clearly observed\cite{Stricker:2014cx} in Sr$_{2}$RuO$_{4}$ at low temperatures.  Despite the clean Fermi liquid behaviour, exemplified in that case by $p=2$, these authors found that at elevated temperatures deviations from the predicted dome shape appeared, which they linked to the increasing importance with increasing temperature of 'resilient' quasi-particles. This observation provides the means to make a direct comparison between the optical conductivity and theoretical calculations, where one does not have to resort to making the decompositions involved in the extended Drude analysis presented in Fig. 1.  To facilitate a direct comparison between experiment and theory we introduce the function $\Delta\sigma(\omega,T)\equiv\sigma_{1}(\omega,T)-\sigma_{2}(\omega,T)$, which is readily obtained from experimental data and also from calculations of the optical conductivity. For the particular case of a local FL, the function $\Delta\sigma(\omega,T)$ has the property that it is negative in the thermal regime where characteristic FL behaviour should be observed, while it is positive in the incoherent and Drude-like regimes.  Moreover, the zeros of this function correspond to the ``dome" derived by Berthod \textit{et al.}.  $\Delta\sigma(\omega,T)$ thus allows us to examine the full, complex optical conductivity and search for zero-crossings where $\sigma_{1}(\omega,T)=\sigma_{2}(\omega,T)$.

In Fig. 2{\bf a}, $\Delta\sigma(\omega,T)$ is displayed as a false color plot for the annealed crystal. In Fig. 2{\bf a}, blue represents $\Delta\sigma(\omega,T)<0$, while red indicates $\Delta\sigma(\omega,T)>0$.  $\Delta\sigma(\omega,T)=0$ is indicated in white. The most striking feature of Fig. 2{\bf a} is a clear dome of zero crossings, closely resembling the dome predicted by Berthod \textit{et al.}. In Fig. 2{\bf b} we display calculations of $\Delta\sigma(\omega,T)$, assuming a FL self-energy (see Methods and SOM\cite{som_annpnic} for calculation details).  Motivated by the saturation of $M_{2}(\omega,T)$ above 50 meV (Fig 1{\bf d}), we have introduced a cutoff $\omega_{c}$ above which the imaginary part of the single particle self-energy, $\Sigma_{2}(\omega,T)$, is constant (see Fig. 3{\bf a} and Methods) and a high energy cutoff $D$. The calculated $\Delta\sigma(\omega,T)$ is in excellent agreement with the experimental data. As input for the calculation we have used several experimentally available parameters, namely $\omega_{p}\approx$ 1.4 eV, $\Gamma_{0}=M_{2}(\xi\rightarrow 0)\approx$ 7 meV and $p\approx$ 1.47. The cutoffs $\omega_{c}\approx$ 41 meV and $D\approx$ 1 eV are motivated below. In addition to the free charge response, we also include the frequency dependent interband response from Supplementary Table S1\cite{som_annpnic}. The only remaining free parameter, $T_{0}\approx$ 1700 K, is determined by two criteria: (i) the maximum of the dome of zero crossings (at $\hbar\omega\approx$ 55 meV) and (ii) the low temperature zero-crossing at $\hbar\omega\approx$ 100 meV. To facilitate the estimation of $T_{0}$, we derive an approximate analytical expression, $T_{\infty}(\omega)$ (see Methods and SOM\cite{som_annpnic}), for these zero-crossings taking an energy independent interband response ( e.g.  $\varepsilon_{\infty}$) into account. The consistency between $T_{\infty}(\omega)$, the data, and the calculation (which includes the full frequency dependence of the interband response) shows that the details of the interband response are unimportant for obtaining this level of agreement.
 
\subsection{Characteristic Fermi liquid properties of Co-doped BaFe$_{2}$As$_{2}$.\\}
The deviation from scaling in Fig. 1{\bf e-g} around 100 K signals a crossover temperature where $\hbar\omega\leq p\pi k_{B}T$, above which an incoherent regime emerges\cite{Maslov:2012ba,Berthod:2013qq}. This suggests a natural cutoff $\hbar\omega_{c}\approx$ 1.47$\pi k_{B}T$ with $T\approx$ 100 K, resulting in $\omega_{c}\approx$ 41 meV. The cutoff $D\approx$ 1 eV is less critical but is motivated by the value of $T_{0}$. Dynamical Mean Field Theory (DMFT) calculations for a single band Hubbard model\cite{Berthod:2013qq} indicate that $k_{B}T_{0}\approx 0.57\delta W$ where $W$ is half the bandwidth and $\delta$ is the carrier density. This yields $W\approx$ 1.3 eV in our case, which is reasonable compared to combined density functional theory and DMFT (e.g LDA+DMFT) estimates\cite{werner:2012cs}. 

Apart from the saturation in $\Sigma_{2}(\omega,T)$, $\omega_{c}$ also introduces\cite{som_annpnic} a frequency dependence in $\Sigma_{1}(\omega,T)$ (see Fig. 3{\bf a}), which should be manifest as a frequency dependent mass enhancement $m^{*}/m(\omega,T)\equiv 1+M_{1}(\omega,T)/\omega$ in the free charge response. Fig. 3{\bf b} shows excellent agreement between $m^{*}/m(\omega,T)$ extracted from experiment and the theoretical calculation where $m^{*}/m(\omega\rightarrow 0,T)\approx$ 1.2. This value is consistent with a modest $m^{*}/m\approx$ 1.8 predicted by LDA+DMFT calculations for this level of electron doping.\cite{werner:2012cs} The experimental data leaves some room for additional mass enhancement resulting from boson exchange processes (such as phonons or spin-fluctuations) below $\hbar\omega\approx$ 10 meV, although it is difficult to make a quantitative statement on their strength due to the low signal-to-noise at low energy. More importantly, the energy dependence of the mass-enhancement introduced through the cutoff in our self-energy rules out the presence of a significant boson exchange spectrum for $\hbar\omega\geq$ 10 meV. We emphasise that the calculated mass enhancement is based on an analysis of the optical conductivity, while the experimental mass enhancement is determined using the extended Drude analysis presented in Fig.  1. The excellent agreement between the experimental and calculated $m^{*}/m(\omega,T)$ therefore serves as a confirmation of the analysis presented in Fig. 1. 

\section{Discussion}
To conclude we discuss the deviation of $p$ from the FL value $p$ = 2. The most likely origin appears to be scattering of quasi-particles on weak, localised magnetic moments\cite{Maslov:2012ba}. Such localised moments could be associated with the presence of Co impurities in the Fe lattice, although no local moment has been detected for Co impurities in BaFe$_{2}$As$_{2}$\cite{Ning:2008ei}. Regardless the origin, this resonant elastic term has a strong influence on the normal state properties. Figure 3c displays $\Delta\sigma(\omega,T)$ for the as-grown crystal, displaying a suppressed dome of zero-crossings compared to the annealed crystal. The dashed semi-circle is calculated using exactly the same parameters as for the annealed case, except for a slightly higher $\Gamma_{0}\approx$ 8 and $p$ = 1.34. This smaller value of $p$ corresponds to a two-fold stronger elastic term in the single particle self-energy $\Sigma(\omega,T)$, indicating that annealing strongly reduces the influence of this scattering channel. Given the concomitant change in superconducting critical temperature, we suggest that this scattering channel could be pair-breaking, possibly providing an interesting direction for future work.

\subsection{Methods.}
A large 4 x 5 x 0.1 mm$^{3}$ single crystal of BaFe$_{1.8}$Co$_{0.2}$As$_{2}$, grown from self-flux, was cut into two pieces and one piece was subsequently annealed for 75 hours at 800 $\degree$C. The dc resistivity and dc susceptibility show an increase of the critical temperature $\Delta T_{c}/T_{c}\approx$ 0.3 upon annealing, while the overall value of the resistivity decreases. Further details of the experiments are presented in the SOM \cite{som_annpnic}.
The theoretical formalism is based on the Allen-Kubo formula for the optical conductivity,\cite{Allen:PRB1971}
\begin{equation}\label{Kubo}
\sigma(\omega ,T) = \frac{{\omega _p^2 }}{{i4\pi\omega}}\int\limits_{ -
\infty }^{ + \infty }  \frac{{n_{F}(\omega + x,T) - n_{F}(x,T)}}{{\omega  - \Sigma (x + \omega ,T)
+ \Sigma ^* (x,T) + i\Gamma_{imp}}}dx
\end{equation}
which we evaluated numerically. The imaginary part of the single particle self-energy appearing in the denominator is given by,\cite{Maslov:2012ba}
\begin{equation}\label{SEFL}
\Sigma_{2}(\omega,T)=-\frac{i}{\pi k_{B}T_{0}}\left[(1+a)(\hbar\omega)^2+(\pi k_{B}T)^{2}\right]
\end{equation}
for a local Fermi liquid with an additional elastic resonant scattering contribution (note $a=(p^{2}-4)/(1-p^{2})$\cite{Maslov:2012ba}). Such an energy and temperature dependent $\Sigma_{2}(\omega,T)$ results at low temperature in an imaginary memory function,\cite{Maslov:2012ba, Berthod:2013qq}
\begin{equation}\label{MFL2}
M_{2}(\omega,T)=\frac{2}{3\pi k_BT_{0}}\left[(1+a)(\hbar\omega)^{2}+(2\pi k_{B}T)^{2}\right]
\end{equation}  
Together with equation (\ref{EDM}) for the optical conductivity and an interband contribution characterised by an energy independent value $\varepsilon_{\infty,IR}$, equation (\ref{MFL2}) leads to the following expression for the dashed semi-circle displayed in Fig. 2:
 \begin{widetext}
\begin{equation}
T_{\infty}(\omega)=\frac{1}{4}\sqrt{\frac{3}{\varepsilon_{\infty,IR}\pi\omega}\sqrt{T_{0}^{2}\left[\omega_{p}^{4}+4\varepsilon_{\infty,IR}\omega^{2}\omega_{p}^{2}-4\varepsilon^{2}_{\infty,IR}\omega^{4}\right]}-\frac{4\omega^{2}(1+a)}{\pi^{2}}-\frac{3T_{0}\omega^{2}_{p}}{\varepsilon_{\infty,IR}\pi\omega}-\frac{6\Gamma_{0}T_{0}}{\pi}}.
\end{equation}
 \end{widetext}
For a derivation and further details see the SOM \cite{som_annpnic}. The parameters used to calculate the dashed semi-circles in Fig. 2{\bf a,b} are the same as for the full calculation except for $\varepsilon_{\infty,IR}\approx$ 100. For the full calculation of equation (\ref{Kubo}), we introduce two cutoff's, $\omega_{c}$ and $D$ in equation (\ref{SEFL}) such that the imaginary part of the self-energy is given by $\Sigma_2(\omega) \propto \omega^2$ for $|\omega| < \omega_c$; $\Sigma_2(\omega) \propto \omega_c^2$ for $|\omega| \in [\omega_c,D]$; and $\Sigma_2(\omega) = 0$ otherwise. In the SOM \cite{som_annpnic} we derive analytical expressions for the real part of the self-energy obtained from Kramers-Kronig transformation.

\section{Acknowledgements}
E.v.H would like to acknowledge stimulating discussions with A.V. Chubukov, I. Eremin, B. B{\"u}chner and in particular C. Berthod. E.v.H would also like to thank H. Luigjes, T. de Goede and K. de Nijs for experimental support.
A.T. carried out and analysed experiments. As-grown and annealed single crystals were provided by Y.H.  A.T. and A.d.V. carried out and analysed transport and magnetisation experiments on the crystals used in this study. S.J. performed theoretical calculations, provided theoretical support and wrote the manuscript. E.v.H designed the experiment, analysed data, performed theoretical calculations and modelling, and wrote the manuscript.
Correspondence and requests for additional material should be addressed to E.v.H (email: e.vanheumen@uva.nl).


\newpage

\renewcommand\thefigure{S\arabic{figure}} 
\setcounter{figure}{0}
\renewcommand{\thetable}{S\arabic{table}}
\renewcommand{\theequation}{S\arabic{equation}}
\setcounter{equation}{0}
\section{Supplementary material}
\section{Transport experiments}
\begin{figure*}[tbh]
\centering
\includegraphics[ height = 6 cm]{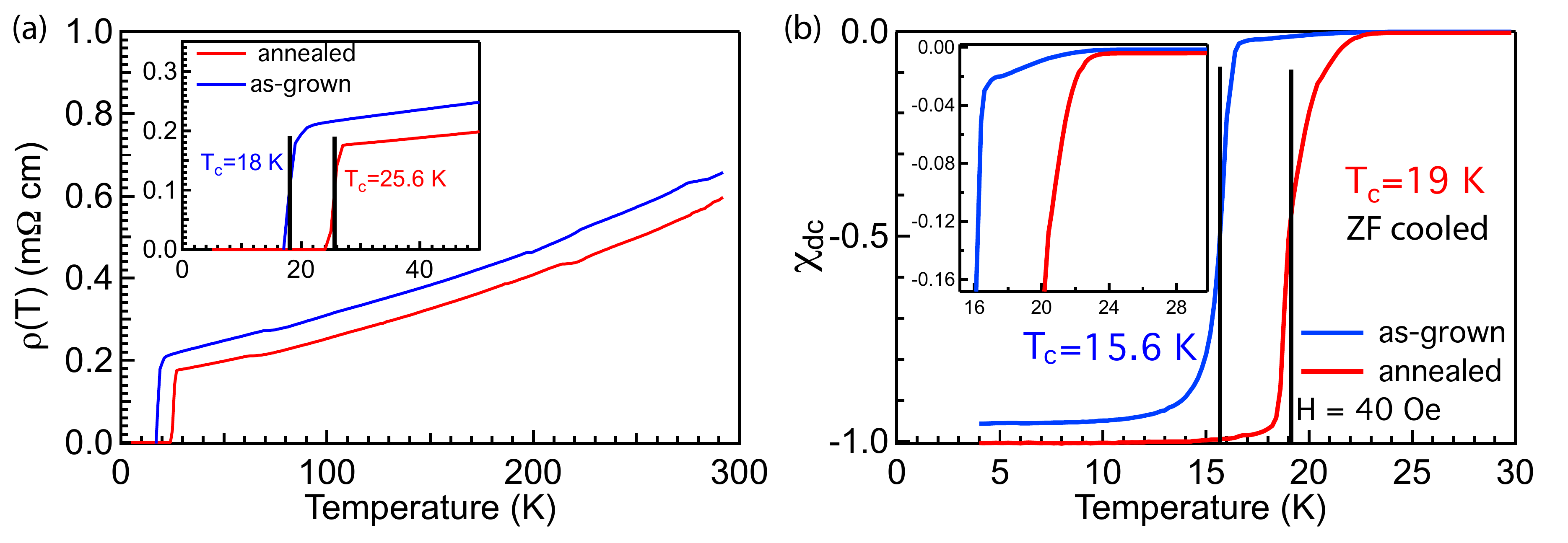}
\caption{(Color online) \textbf{Transport and magnetisation measurements.} Two crystals were measured: as-grown (blue) and annealed (red) crystals. (a) dc resistivity with an approximate T$^{2}$ dependence up to 300 K. The little kinks at 70 K and 220 K are experimental artefacts. The inset shows an enlarged view of the superconducting transitions to the zero-resistance state. (b) dc susceptibility measured in a 40 Oersted field. As the inset shows the onset of the transition takes place at significantly higher temperature compared to the temperature where the full volume of the crystal becomes superconducting.}
\label{S1}
\end{figure*}

The crystal used in this study was grown from a self-flux method. Its chemical composition has been determined with electron probe microanalysis, resulting in the determination of a Co concentration, $x$ = 0.195. We subsequently cut the crystal into two pieces and annealed one piece for 75 hours at 800 $\degree$C. The dc resistivity and susceptibility were measured for both pieces, see Fig. \ref{S1}.
The resistivity shows a significant enhancement of the critical temperature from 18 K for the as-grown crystal to 25.6 K in the annealed crystal. The dc susceptibility shows an onset to superconductivity at similar temperatures, but the full Meisner volume is obtained at somewhat lower temperatures. 

\section{Reflectivity data}
\begin{figure*}[tbh]
\includegraphics[ width = 12 cm]{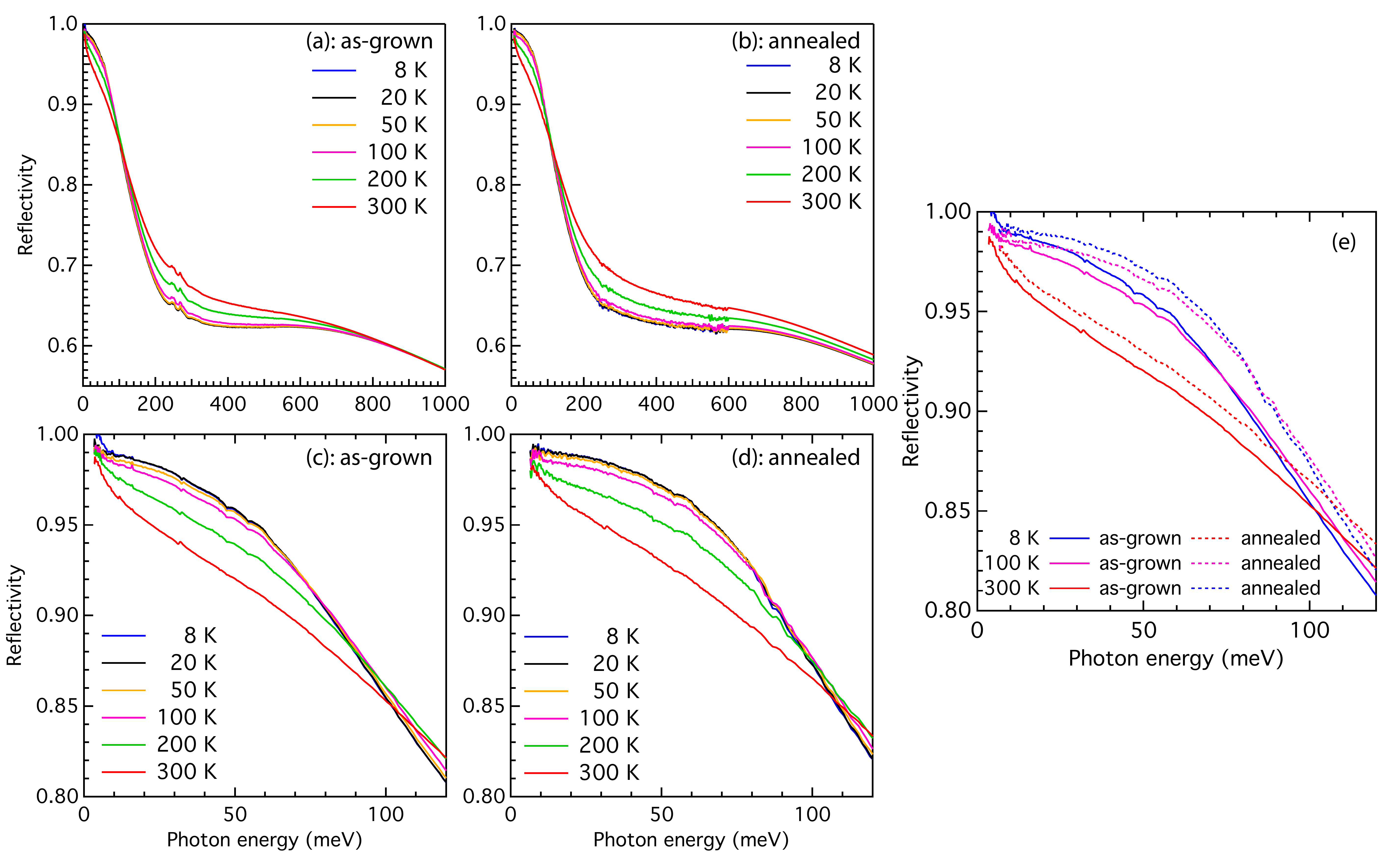}
\caption{(Color online) \textbf{Reflectivity experiments.} Reflectivity of the as-grown (a,c) and annealed (b,d) crystals at selected temperatures.}
\label{S2}
\end{figure*}

The in-plane reflectivity for both crystals was obtained in the range between 5 meV and 4.6 eV using a Bruker vertex 80v Fourier-Transform infrared spectrometer. In all experiments unpolarised light was used and experiments were performed at near-normal incidence (8 degrees) to the ab-plane. Crystals were mounted on a tapered copper cone, preventing unwanted reflections from copper, and were cleaved before inserting them into a home-built UHV cryostat. The cryostat has a rigid sample support decoupled from the cryostat cold finger with a copper braid. This enables optical experiments as function of temperature in which the sample position does not change as temperature is changed. To test sample position stability of the cryostat a HeNe laser was reflected from a mirror mounted on the sample position and the movement of the reflected beam over a distance of 5 meters was measured, confirming that the sample orientation changed less than 0.01 $\degree$ between room temperature and the base temperature of 8 K. In the experiments an aperture size was chosen such that the light spot slightly overfills the sample surface. All experiments were performed under UHV conditions with a pressure at room temperature of order 5$\cdot$ 10$^{-9}$ mbar. A wedged, CVD grown diamond window was used in all experiments. To cover the full energy range, experiments were repeated several times using a series of detectors and beamsplitters. In order to obtain an accurate absolute value of the reflectivity we evaporated metallic films on our samples \text-it{in-situ}. Au was used in the far to mid-infrared range (3 meV - 0.75 eV). Silver was used in the mid-infrared to visible range (0.4 - 2.85 eV) and Al in the near-infrared to ultra-violet range (0.75 - 4.6 eV). In the near-infrared to visible range measurements were performed in 3 steps. In the first step the reflection of the sample surface was measured. In the second step Ag was evaporated \textit{in-situ} on the sample surface without making any adjustments to the set-up. Finally in the third step Al was evaporated \textit{in-situ} on the sample surface. For each of these steps the full temperature dependence was measured by collecting 1 spectrum per minute while cooling down with a constant rate of 1.5 K per minute, followed by a similar warming measurement. By combining both warming and cooling measurements we thus obtained 1 spectrum every 2 K. By comparing ratios of Ag and Al spectra to published literature results we obtained standardized reference spectra as function of temperature used to determine the reflectivity in the entire measured frequency range. Silver turned out to be particularly useful in the photon range around 1.2 eV where Al has a weakly temperature dependent interband transition.
The resulting reflectivity is shown in figure \ref{S2} for selected temperatures.  
For the as-grown crystal we observe the opening of the superconducting gap below 18 K as can be seen in panel \ref{S2}c. The somewhat smaller crystal size (2 x 1 x 0.1 mm$^{3}$) for the annealed crystal complicated the accurate determination of the reflectivity below 10 meV most likely due to diffraction effects becoming important at these longer wavelengths. Figure \ref{S2}e compares the low frequency reflectivity of the annealed and as-grown crystals. Based on the higher reflectivity of the annealed crystal one can immediately observe that the overall scattering rate has decreased, assuming that the charge carrier concentration has not significantly changed. 
In a first step the reflectivity data is modelled using a Drude-Lorentz model with parameters optimized by a least-square Levenberg-Marquardt routine \cite{kuzmenko:reffit}. We tried different models to determine the robustness of the modelling, as we want to use it later on in the extended Drude analysis. The Drude-Lorentz models presented in table \ref{DLtable} for both crystals, (i) give the lowest $\chi^{2}$, (ii) consistently describe our optical data at all temperatures and (iii) are nearly identical for both crystals as one might expect. Based on this model and the full reflectivity data we use a variational dielectric function routine developed in \cite{Kuzmenko:2005jhSP} to extract the optical conductivity, which is shown at selected temperatures in Fig. 1a,b of the article. 

\section{Drude-Lorentz model and interband transitions}\label{interbandsec}
\begin{figure}[tbh]
\centering
\includegraphics[ width = 8.6 cm]{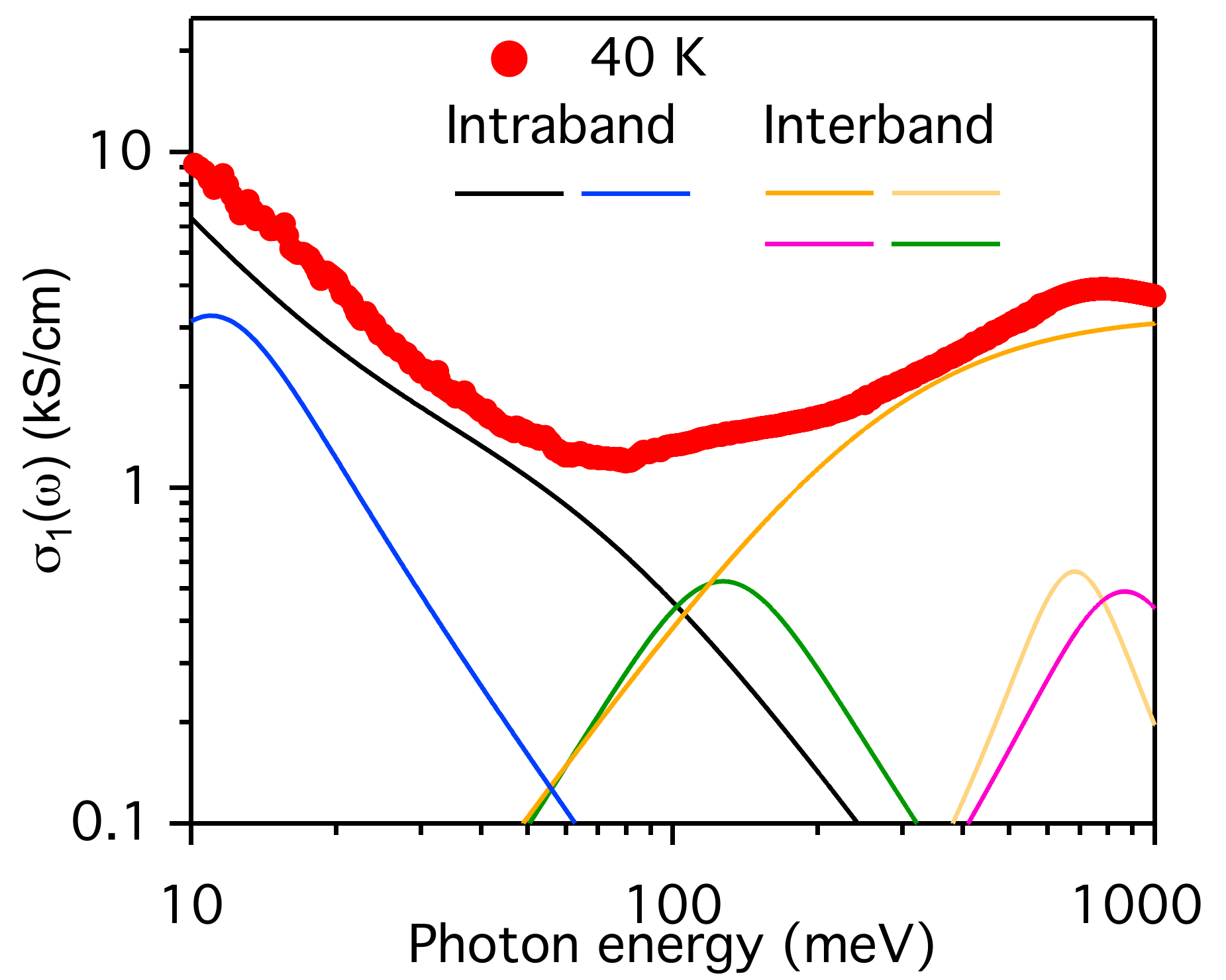}
\caption{(Color online) \textbf{Decomposition of the optical conductivity.} Experimental spectrum for the annealed crystal at 40 K (symbols) and the decomposition in oscillators (solid lines) corresponding to the parameters given in table \ref{DLtable}. Note that the two Drude components are plotted as a single curve (black).}
\label{S3}
\end{figure}

\begin{table}
\centering
\caption{\textbf{Parameters of the Drude-Lorentz oscillators.} The high-frequency dielectric constant $\epsilon_{\infty}\approx$ 9. $\omega_j$ is the centre frequency of an oscillator, $\omega_{pj}$ its area (Drude terms) or oscillator strength and $\gamma_j$ is the width. All values reported correspond to T 40 $K$ data.}
 \label{DLtable}
  \begin{tabular}{|c|*{8}{c}}
as-grown\\
  \hline
j & 1	& 2 & 3 &	4 & 5	& 6 & 7 &	8 	 \\
\hline 
$\hbar\omega_j$ (eV) &0 &0 &0.01 &0.12 &0.67 &0.92 &1.6 &1.82 \\
$\hbar\omega_{pj}$ (eV) &1 &0.96 &0.42 &0.59 &2.49 &1.96 &14.56 &3.65 \\
$\hbar\gamma_j$ (eV) &0.005 &0.065 &0.01 &0.12 &0.58 &0.73 &14.55 &1.84\\
 \hline
annealed\\
\hline
$\hbar\omega_j$ (eV) &0 &0 &0.011 &0.13 &0.68 &0.87 &1.6 & \\
$\hbar\omega_{pj}$ (eV)   &1.1 &0.8 &0.51 &0.72 &1.28 &1.62 &14.95 &  \\
$\hbar\gamma_j$ (eV) &0.0034 &0.064 &0.01 &0.13 &0.39 &0.72 &9.48 & 
  \end{tabular}
\end{table}
In order to make the spectroscopic fingerprints of the effects of annealing more quantitative we first turn to a standard Drude-Lorentz modelling of the data. The decomposition of the optical conductivity of both as-grown and annealed crystals in Drude and Lorentz terms is given in table \ref{DLtable}. We find that the intraband contribution can be described by two relatively narrow Drude terms and a low energy Lorentz oscillator. The incoherent background extending to low energy is captured by a high energy oscillator (labeled as nr. 7 in table \ref{DLtable}). Upon annealing the width of this oscillator decreases, resulting in a much weaker contribution at low energy. This effect can be clearly seen in the optical conductivity data by comparing the depth of the minimum (at 70 meV) separating the intra- and interband response. Apart from this difference in the incoherent background the Drude-Lorentz models are nearly identical, indicating that the annealing results in a rather subtle change in the intraband response.
The optical conductivity of the annealed crystal at 40 K and its decomposition in terms of the oscillators from Table \ref{DLtable} is shown in Fig. \ref{S3}.  

\section{Spectral weight analysis}\label{SWsec}
\begin{figure*}[tbh]
\includegraphics[ width = 15 cm]{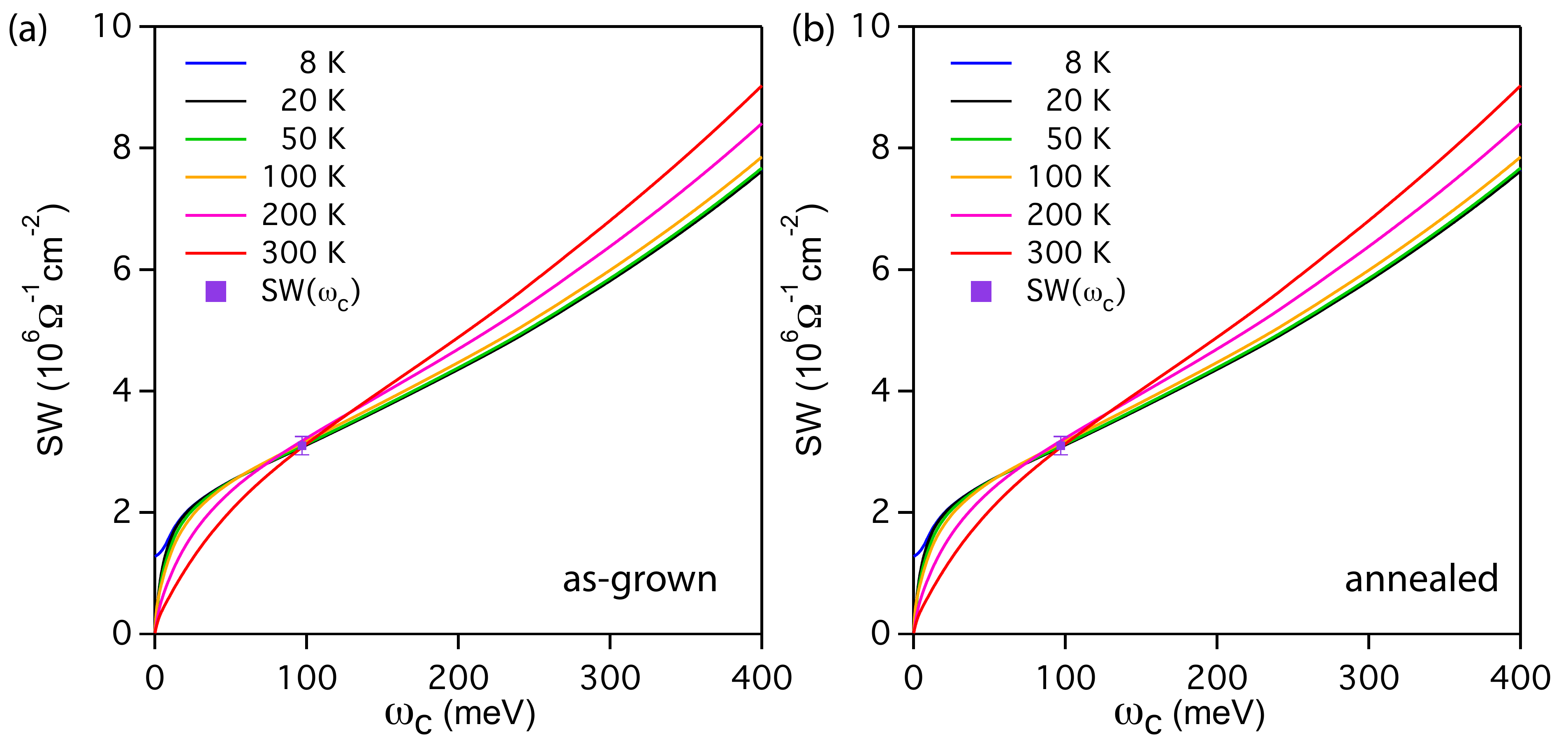}
\caption{(Color online) \textbf{Spectral weight analysis.} Spectral weight of the as-grown (a) and annealed (b) crystals at selected temperatures.}
\label{S4}
\end{figure*}

In order to determine the plasma frequency for both crystals we plot in figure \ref{S4} the spectral weight as determined from the experimental optical conductivity.
The spectral weight is obtained by integrating the real part of the optical conductivityover frequency up to a cutoff frequency $\omega_{c}$:
\begin{equation}
SW(\omega_{c},T)=\int_{0}^{\omega_{c}}\sigma_{1}(\omega,T)d\omega
\end{equation}
For both crystals we find that the integrated spectral weight is nearly temperature independent for $\omega_{c}\approx$ 100 meV, corresponding roughly to the minimum in the optical conductivity presented in Fig. 1a,b of the main text. At this point the integrated spectral weight $SW(\omega_{c},T)\approx$ 3.1$\pm$0.15 $\cdot$ 10$^{6}$ $\Omega^{-1}cm^{-2}$ (as-grown) and $SW(\omega_{c},T)\approx$ 3.4$\pm$0.15 $\cdot$ 10$^{6}$ $\Omega^{-1}cm^{-2}$ (annealed). If we assign this spectral weight in both cases entirely to the intraband response we can calculate the plasma frequencies to be $\omega_{p}\approx $ 1.35 eV (as-grown) and $\omega_{p}\approx $ 1.4 eV (annealed). Given the error bar on the estimation of $SW(\omega_{c},T)$ we use $\omega_{p}\approx $ 1.4 eVfor both crystals.

\section{Extended Drude analysis: interband contributions and range of validity.}
\begin{figure}[tbh]
\centering
\includegraphics[ width = 8.6 cm]{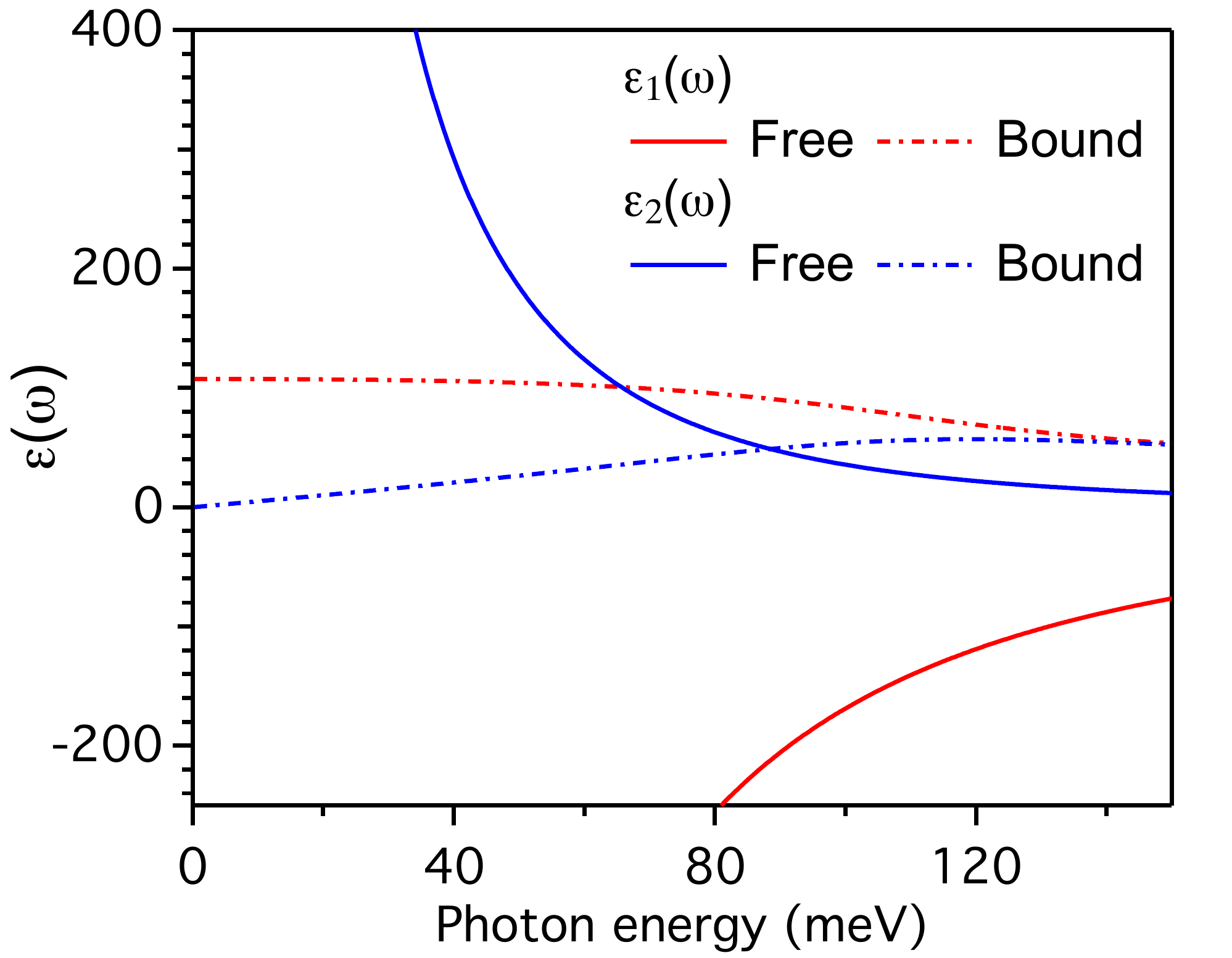}
\caption{(Color online) \textbf{Dielectric response.} Decomposition of the dielectric function in bound and free charge response at 40 K corresponding to the model parameters of table \ref{DLtable}.}
\label{S5}
\end{figure}
\begin{figure*}[tbh]
\centering
\includegraphics[ width = 14 cm]{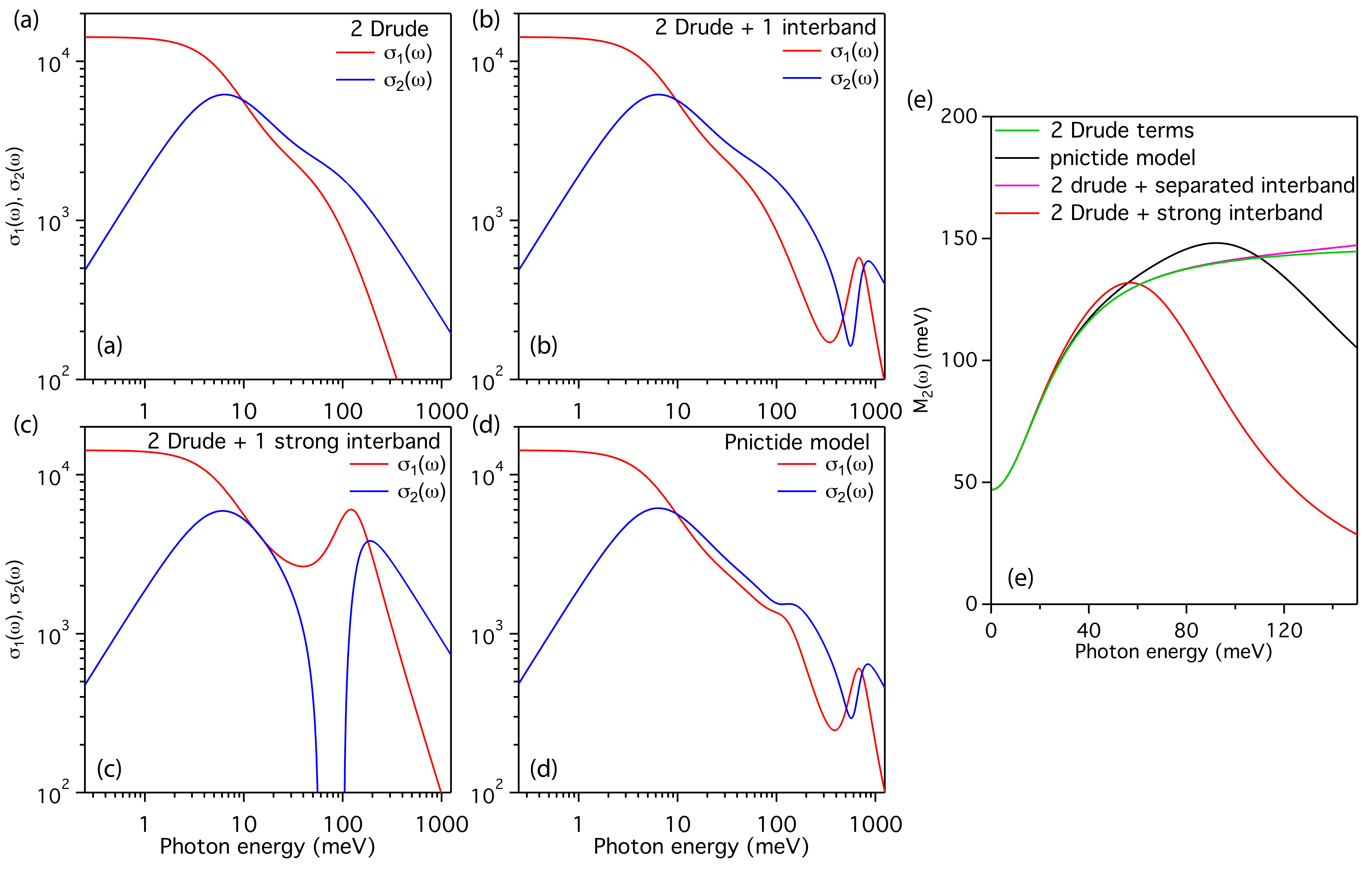}
\caption{(Color online) Model optical conductivity for (a): superposition of two Drude terms, (b): two Drude terms and a well separated interband transition, (c): two Drude terms and a strong interband transition overlapping with the intraband response and (d): a model used for the annealed crystal. (e): Memory functions calculated for each of the panels (a-d) without taking the presence of the interband transitions into account. The comparison shows that the additional frequency dependence of interband processes is modest for photon energies up to 50 meV for the worst case considered (e.g. two Drude terms + overlapping strong interband transition).}
\label{S6}
\end{figure*}
As discussed in the main text, overlapping intra- and interband conductivities complicate the extraction of the memory function from optical conductivity data. In the absence of overlapping intra- and interband conductivities (e.g. such as is the case for cuprate HTSC) one can write  the memory function in terms of the dielectric function:
\begin{eqnarray}\label{ED2}
M_{1}(\omega)=\frac{\omega^{2}_{p}}{\omega}\frac{\varepsilon_{2}(\omega)}{[\varepsilon_{\infty,IR}-\varepsilon_{1}(\omega)]^{2}+\varepsilon_{2}^{2}(\omega)}\\
M_{2}(\omega)=\frac{\omega^{2}_{p}}{\omega}\frac{\varepsilon_{\infty,IR}-\varepsilon_{1}(\omega)}{[\varepsilon_{\infty,IR}-\varepsilon_{1}(\omega)]^{2}+\varepsilon_{2}^{2}(\omega)} -\omega
\end{eqnarray}
where $\varepsilon_{\infty,IR}$ represents a frequency independent contribution to the \textit{real part} of the dielectric function due to high energy interband transitions, which can be estimated from the oscillator strengths of those transitions. What happens in the case of iron-pnictides where interband transitions have a low energy onset (estimated to be situated around 100 meV, see fig. \ref{S3})? This is most clearly illustrated in figure \ref{S5} where we show the free charge (or intraband) and bound charge (or interband) contributions to the dielectric model presented in table \ref{DLtable}. The question that arises is whether the bound charge response can be approximated with a constant in the photon energy range where we want to analyze the memory function. Given the relative strengths of the free and bound charge response, the approximation of using a frequency independent $\varepsilon_{\infty,IR}$ could possibly be upheld below about 50 meV. This is fortified by explicitly calculating the imaginary part of the memory function for a series of models as we will now discuss. 

Fig. \ref{S6} shows the optical conductivity for 4 different models. In panel \ref{S6}a the real and imaginary part of a sum of two Drude terms with different widths is shown. Panel \ref{S6}b,c show the same model but now with a single interband transition added to it. In panel \ref{S6}b the intra- and interband parts are well separated as in the cuprates, while panel \ref{S6}c has a strong interband transition well within the intraband region. Finally, panel \ref{S6}d shows the optical conductivity for a model similar to the pnictide model. Panel \ref{S6}e now compares the extracted memory functions for these models \textit{without} making any correction for the interband contribution. The 2 Drude case (in green) would represent the correct optical scattering rate that we would like to extract in an experiment. The other cases show deviations from this ideal curve to varying degrees. What is important for the current work is that below about 50 - 80 meV the frequency dependence in all cases is very close to the ideal case indicating that in the realistic case relevant to the iron-pnictides (black curve) the extended Drude model gives relevant results in the photon energy range discussed in the main text.   

\section{Extended Drude model: comparison of methods}
\begin{figure}[tbh]
\centering
\includegraphics[ width = 8.6 cm]{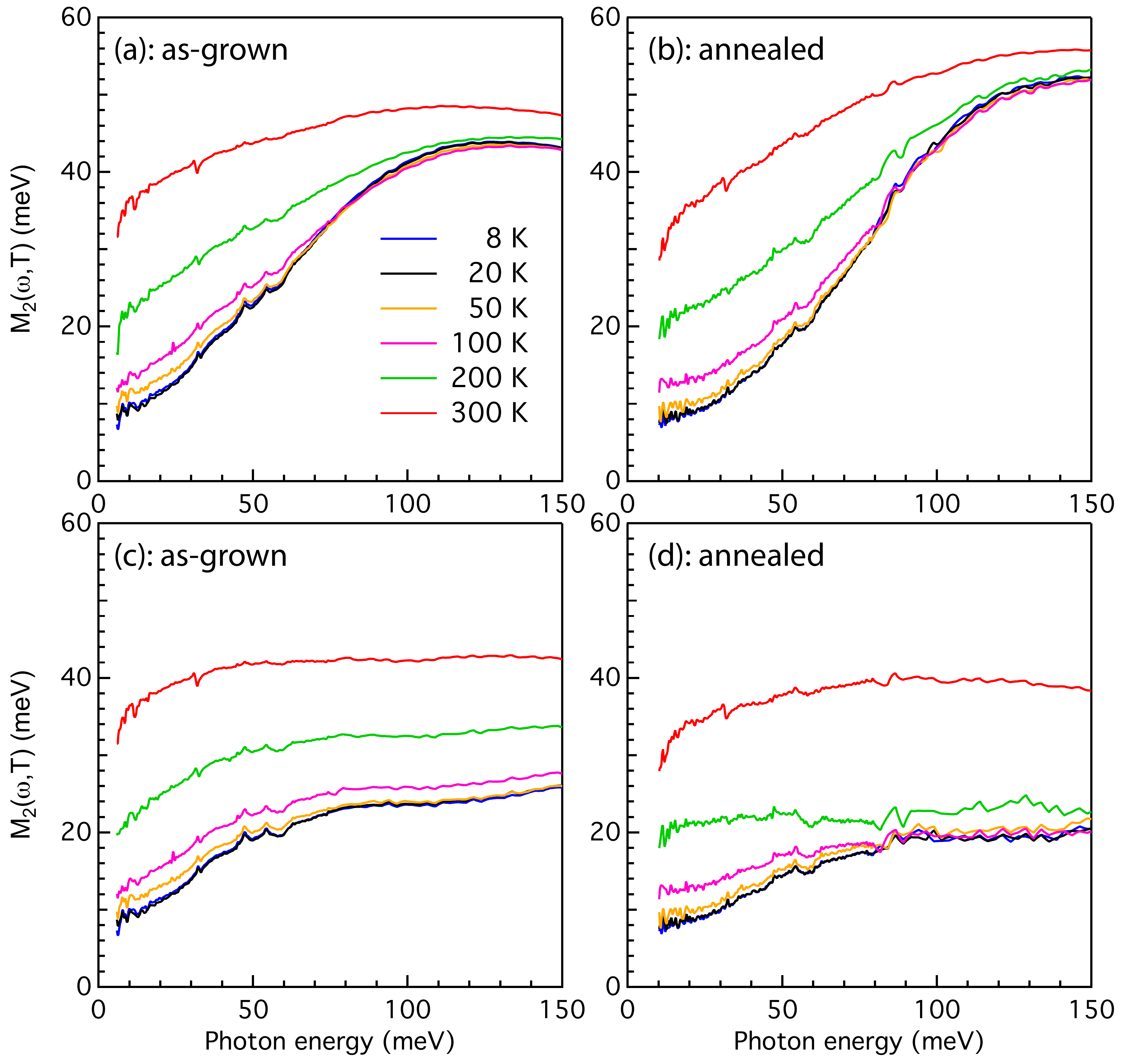}
\caption{(Color online) \textbf{Memory functions.} Comparison between the memory functions obtained using the method of subtracting $\varepsilon_{\infty,IR} \approx$ 100 -110 (a,b) and by subtracting the full frequency dependent interband conductivity (c,d). }
\label{S7}
\end{figure}

\begin{figure}[tbh]
\centering
\includegraphics[ width = 8.6 cm]{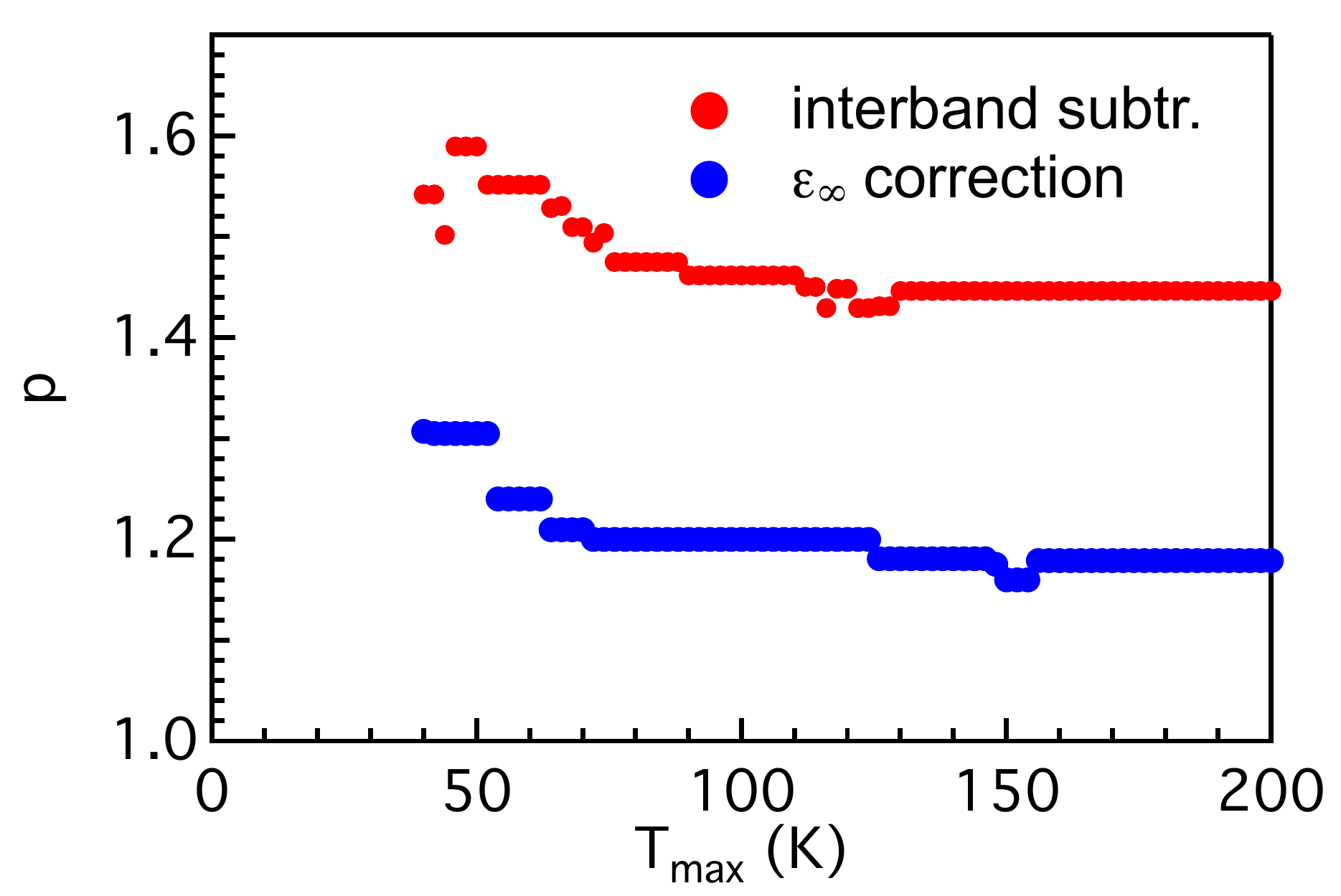}
\caption{(Color online) \textbf{Minimum $p$ values.} Value of $p$ giving the best scaling collapse as a function of maximum temperature used in the determination of the deviation of the data from a universal curve.}
\label{S8}
\end{figure}

To test the robustness of the results presented in the main text we used two methods to determine the memory function. In the first method we approximated the interband contribution with a temperature independent $\varepsilon_{\infty,IR} \approx$ 100 - 105 (for as-grown and annealed crystals respectively). We then used Eq. \ref{ED2} to calculate the memory function. In the second method we subtracted the interband part obtained from the Drude - Lorentz model obtained at each individual temperature. The results of the second method are presented in the main text and the results of the first method are presented in fig. \ref{S7}.
At low temperatures ( $\leq$ 150 K) and photon energies between 10 - 50 meV both methods give nearly identical results. At higher temperatures differences are starting to become more evident as can be seen most clearly by comparing the 200 K data for the annealed crystal. Note that the main result of our article, namely the T$^{2}$ and $\omega^{2}$ dependence of $M_{2}(\omega,T)$, would be extended over a larger energy and temperature range if we use the first method to determine $M_{2}(\omega,T)$ but with a somewhat smaller value of $p$.  
Next we discuss the estimation of the value of $p$ for which all the data collapses onto a universal curve. In ref. \cite{Stricker:2014cxSP} the following method was proposed: one plots the data as function of $\xi^{2}=\left[(\hbar\omega)^{2}+(p\pi k_{B}T)^{2}\right]$ for a range of values of $p$. We take  $1\leq p\leq2$ with steps of 0.01. One then calculates the root-mean square for each value of $p$ determined by summing over the deviations of each temperature from a universal curve for that value of $p$. These RMS values are then summed over a range of temperatures up to a certain maximal temperature. Figure \ref{S8} shows the dependence of p on the maximum temperature, T$_{max}$, used in the scaling analysis. We apply this method to the memory function extracted with both methods indicated above and find that the value of $p$ depends weakly on temperature. At 100 K, where the power of the frequency dependence starts to deviate from $\eta\approx$ 2 we find the values $p$ = 1.2 ($\varepsilon_{\infty}$ correction) and $p$ = 1.47 (interband subtraction). 

\begin{figure*}[tbh]
\centering
\includegraphics[ width = 12 cm]{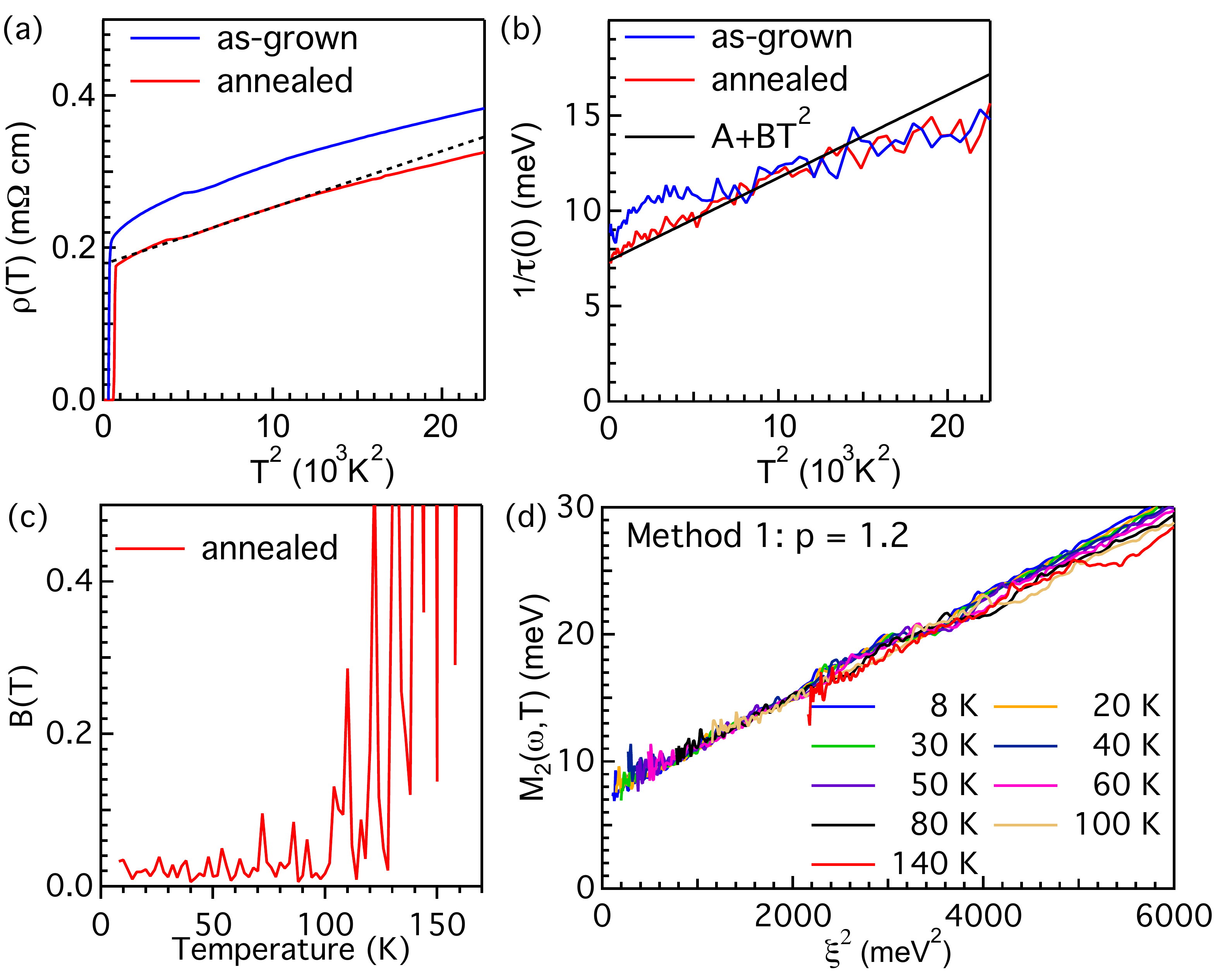}
\caption{(Color online) \textbf{Scaling analysis.} (a) Resistivity as function of $T^{2}$. The dashed line is a guide to the eye. (b): DC scattering rate as function of $T^{2}$. This panel is equivalent to Fig. 1e of the main text (c): Prefactor, B(T), of the frequency component of the memory function. (d): Scaling collapse of $M_{2}(\omega,T)$ obtained by correcting $M_{2}$ with a frequency independent interband contribution.}
\label{S9}
\end{figure*}

In the main article we showed that the optical data can be scaled as a function of $\omega^{2}$ and $T^{2}$. In particular, we showed that the DC extrapolation of the scattering rate follows a $T^{2}$ temperature dependence, implying that the resistivity is also a function of $T^{2}$. To confirm this expectation we show in figure \ref{S9}a the resistivity data of fig. \ref{S1} plotted as function of $T^{2}$. We indeed find that $\rho(T)$ is an approximate function of $T^{2}$ in the same range of temperatures ( 30$\leq$ T $\leq$120 K) as $1/\tau(0)$ (see fig. \ref{S9}b). From fig. \ref{S9}a we estimate $d\rho/d(T^{2})\approx$ 6.5$\cdot10^{-9}$ $\Omega cm/K^{2}$, while $d(\tau)^{-1}/d(T^{2})\approx$ 3.5$\cdot10^{-3}$ $ cm^{-1}/K^{2}$. 
We can now use the Drude expression for the DC resistivity $\rho=4\pi/\omega_{p}^{2}\tau$ to compare the slopes of both quantities. Together with the plasma frequency $\omega_{p}\approx$ 11290 $cm^{-1}$ we obtain $d\rho/d(T^{2})\approx$ 1.7$\cdot10^{-9}$ $\Omega cm/K^{2}$ from $d(\tau)^{-1}/d(T^{2})$. Given the uncertainties involved in determining $1/\tau(0)$ this is a reasonable agreement.
We note that the resistivity deviates from the approximate $T^{2}$ behaviour at the onset of superconductivity and at elevated temperatures. Above the temperature scale where Fermi liquid scaling applies both the DC resistivity and scattering rate are still approximate functions of $T^{2}$, but with slightly smaller slopes as can be seen from the deviation from the black lines.
  
Figure \ref{S9}(c) we show the temperature dependence of the prefactor, B(T), of the frequency component appearing in $M_{2}(\omega,T)=1/\tau(0,T)+B(T)\omega^{\eta(T)}$. Comparing this empirical relation with Eq. 2 of the main manuscript, B(T) is expected to be temperature independent in the range of validity of Eq. 2. Fig. \ref{S9}(c) shows that this is indeed the case below $T\approx$ 100 K. In panel \ref{S9}(d) we plot the scaling collapse for the memory function obtained by subtracting $\varepsilon_{\infty,IR}$. Note that in this case the scaling extends over a larger energy window due to the higher energy where the memory function saturates ($\approx$ 120 meV, fig. \ref{S7}b).  

\section{Zero crossings of $\Delta\sigma(\omega)$ in the presence of interband transitions}
In reference \cite{berthod:2013qqSP} it was shown that in a Fermi liquid the optical conductivity could be described by three different frequency regimes. These regimes are, at low temperature, separated by crossings of $\sigma_{1}(\omega)$ and $\sigma_{2}(\omega)$. It was shown that at low frequency the optical conductivity follows a Drude behaviour, with $\sigma_{1}(\omega)\leq\sigma_{2}(\omega)$, at intermediate frequency a thermal regime appears with $\sigma_{1}(\omega)\geq\sigma_{2}(\omega)$ and finally at high frequency $\sigma_{1}(\omega)\leq\sigma_{2}(\omega)$ again. The boundary separating these regimes can be easily derived from the optical conductivity. We have:
\begin{equation}\label{optcond}
\sigma(\omega)=\frac{i\omega_{p}^{2}}{4\pi}\frac{1}{M(\omega)+\omega}
\end{equation}
with $M(\omega)$ given by \cite{berthod:2013qqSP},
\begin{equation}
M(\omega) = \frac{i}{\pi k_{B}T_{0}}\left[\omega^{2}+(2\pi k_{B}T)^{2} \right].
\end{equation}
By equating $\sigma_{1}(\omega)=\sigma_{2}(\omega)$ one obtains a second order equation relating temperature $T$ and frequency $\omega$. Solving for $T$ gives\cite{berthod:2013qqSP}:
\begin{equation}\label{T1}
T_{1}(\omega)=\sqrt{\frac{3k_{B}T_{0}}{8\pi}\left(\omega-\frac{2\omega^{2}}{3\pi k_{B}T_{0}}\right)}.
\end{equation}
As was shown in ref. \cite{Maslov:2012baSP}, the pre-factor of the temperature term can be different from 2 if additional contributions to the frequency dependence of the self-energy are present, in which case:
\begin{equation}
\Sigma(\omega) = \frac{i}{\pi k_{B}T_{0}}\left[(1+a)\omega^{2}+(\pi k_{B}T)^{2} \right].
\end{equation}
Given the value of $p=1.47$ obtained in our study, we should therefore take,
\begin{equation}
T_{1}(\omega)=\sqrt{\frac{3k_{B}T_{0}}{8\pi}\left(\omega-\frac{2(1+a)\omega^{2}}{3\pi k_{B}T_{0}}\right)}.
\end{equation}
with $a=(p^{2}-4)/(1-p^{2})$.  
In the iron-pnictide superconductors a low energy interband contribution to the optical conductivity further complicates matters. In the article we therefore use the full Allen-Kubo formula and calculate $\Delta\sigma(\omega,T)$, including the full frequency dependent interband conductivity. It is however instructive, and useful for other materials, to approximate the interband contribution with a purely reactive component and derive an analytic expression for the zero crossings. As shown in the main text the zero crossings of this expression and the full calculation do not differ too much. We therefore start from,
\begin{equation}
\sigma(\omega)=\frac{i\omega_{p}^{2}}{4\pi}\frac{1}{M(\omega)+\omega}-\frac{i\omega\varepsilon_{\infty}}{4\pi}
\end{equation}
which is equivalent to,
\begin{equation}\label{optcondmod}
\sigma(\omega)=\frac{1}{4\pi}\left[\frac{\omega_{p}^{2}M_{2}(\omega)+i\left[\omega_{p}^{2}\bar{\omega}-\omega\varepsilon_{\infty}\left(\bar{\omega}^{2}+(M_{2}(\omega))^{2}\right)\right]}{\bar{\omega}^{2}+(M_{2}(\omega))^{2}}\right]
\end{equation}
with $\bar{\omega}=\omega+M_{1}(\omega)$. 
We consider again the case where $M_{1}(\omega)$=0 and $M_{2}(\omega)\propto\left[(1+a)\omega^{2}+(2\pi k_{B}T)^{2}\right]+\Gamma_{0}$. Here the last term is an additional, frequency independent impurity scattering rate. Solving Eq. \ref{optcondmod} for  $\sigma_{1}(\omega)=\sigma_{2}(\omega)$ results in a fourth-order equation in $T$ and $\omega$, which has only one physical solution:
\begin{widetext}
\begin{equation}
T_{\infty}(\omega)=\frac{1}{4}\sqrt{\frac{3}{\varepsilon_{\infty}\pi\omega}\sqrt{T_{0}^{2}\left[\omega_{p}^{4}+4\varepsilon_{\infty}\omega^{2}\omega_{p}^{2}-4\varepsilon^{2}_{\infty}\omega^{4}\right]}-\frac{4\omega^{2}(1+a)}{\pi^{2}}-\frac{3T_{0}\omega^{2}_{p}}{\varepsilon_{\infty}\pi\omega}-\frac{6\Gamma_{0}T_{0}}{\pi}}.
\end{equation}
 \end{widetext}
Although this expression looks more unwieldy than Eq. \ref{T1}, most of the parameters can be determined independently from each other. The plasma frequency, $\omega_{p}$, can be determined from a spectral weight analysis, while $\varepsilon_{\infty}$ can be estimated from a Drude-Lorentz analysis. $\Gamma_0$ can be determined at low temperature from an extrapolation of $M_{2}(\omega)$ to $\omega$ = 0. This leaves $T_{0}$ and $p$ as free parameters.

\section{Self-energy of a local Fermi liquid with a cutoff.}
The imaginary part of the complex self-energy $\Sigma(\omega,T) = \Sigma_1(\omega,T) + i\Sigma_2(\omega,T)$ 
of a local Fermi liquid is given by \cite{Maslov:2012baSP,berthod:2013qqSP},
 \begin{widetext}
\begin{equation}
 \Sigma_2(\omega,T) = 
\left\{
 \begin{array}{ccc}  \frac{1}{\pi k_BT_0}[(1+a)(\hbar\omega)^2 + (\pi k_BT)^2]& &0<|\omega|<\omega_c\\
   \frac{1}{\pi k_BT_0}[(1+a)(\hbar\omega_c)^2 + (\pi k_BT)^2] & &\omega_c < |\omega| < D\\
    0 & & D < |\omega|
 \end{array}
\right.
,
\end{equation}
 \end{widetext}
where $D$ represents (a fraction of) the total bandwidth. 
An analytical expression for the real part of the self-energy can be obtained by Kramers-Kronig transformation. It is given by 
 \begin{widetext}
\begin{equation}
\Sigma_1(\omega,T) = \frac{(1+a)}{\pi^2 k_BT_0}2\omega\omega_{c}+\frac{(1+a)}{\pi^2k_B T_0}(\omega_{c}^{2}-\omega^{2})\left[\log\left(\frac{|D-\omega|}{|\omega_{c}-\omega|}\right)+\log\left(\frac{|\omega_{c}+\omega|}{|D+\omega|}\right)\right]+\frac{\Sigma_2(\omega,T)}{\pi}\log\left(\frac{|D-\omega|}{|D+\omega|}\right)
\end{equation}
 \end{widetext}
for $0 < |\omega| < \omega_c$,
 \begin{widetext} 
\begin{equation}
\Sigma_1(\omega,T)=\frac{(1+a)}{\pi^2k_BT_0}2\omega\omega_{c}+\frac{(1+a)}{\pi^2k_BT_0}(\omega^{2}-\omega_{c}^{2})\log\left(\frac{|\omega_{c}-\omega|}{|\omega_{c}+\omega|}\right)+\frac{\Sigma_2(\omega_{c},T)}{\pi}\log\left(\frac{|D-\omega|}{|D+\omega|}\right)
\end{equation}
 \end{widetext}
for $\omega_{c}<|\omega|<D$, and
 \begin{widetext} 
\begin{equation}
\Sigma_1(\omega,T)=\frac{(1+a)}{\pi^2k_BT_0}2\omega\omega_{c}+\frac{\Sigma_2(\omega,T)}{\pi}\log\left(\frac{|\omega_{c}-\omega|}{|\omega_{c}+\omega|}\right)+\frac{\Sigma_2(\omega_{c},T)}{\pi}\left[\log\left(\frac{|D-\omega|}{|\omega_{c}-\omega|}\right)+\log\left(\frac{|\omega_{c}+\omega|}{|D+\omega|}\right)\right]
\end{equation}
 \end{widetext}
for $\omega>D$. 
The real and imaginary parts of this self-energy are plotted in Fig. 3a of the main text.

\end{document}